\documentclass[11pt, oneside]{article}
\usepackage[a4paper, total={6.2in, 9in}]{geometry}
\usepackage{url}
\usepackage{graphicx}
\usepackage{float}
\usepackage{geometry}
\graphicspath{{./}}

\title{Factors in the Portability of Tokenized Assets on Distributed Ledgers}
\author{Richard Barnes \\ Corporate and Investment Bank CTO Office \\ Barclays}
\begin{document}
\maketitle
\begin{abstract}
The tokenization of assets deployed to distributed ledger technology is increasingly cited to revolutionize financial services by allowing traditionally illiquid assets to be bought and sold on primary and secondary markets increasing asset liquidity, 
transparency and reducing transaction completion time. To realize these benefits it is important the token is transferrable, that is, portable from one distributed ledger to another.
In this paper we survey current interoperability architectures and smart contract languages, 
identifying factors affecting the portability of tokenized assets. We propose a portability maturity model that can be used to help assess the current state of technology and supporting market infrastructure.
\end{abstract}

\makeatletter{\renewcommand*{\@makefnmark}{}
\footnotetext{\scriptsize{\textcopyright Barclays Bank PLC 2020
\\
This work is licensed under a Creative Commons Attribution 4.0 International License (CC BY).
\\
Provided you adhere to the CC BY license, including as to attribution, you are free to copy and redistribute this work in any medium or format and remix, transform, and build upon the work for any purpose, even commercially.
\\
BARCLAYS is a registered trade mark of Barclays Bank PLC, all rights are reserved.}
}\makeatother}

\section{\label{sec:introduction}Introduction}
The tokenization of existing physical assets is expected to increase the liquidity of illiquid assets, broaden market accessibility, improve transparency, reduce transaction cost and completion time \cite{deloitte:1}. 
Tokenization refers to the process of digitally representing an existing physical off-chain asset deployed to a distributed ledger \cite{EY:2}. Financial regulators are only now starting to publish guidance frameworks that include definitions of different types of token and the resulting regulation regimes they fall under; this has been mainly in response to Initial Coin Offerings (ICO). 
The Swiss Financial Market Supervisory Authority (FINMA) have categorized tokens into three broad categories: \textit{Payment tokens} which are synonymous with cryptocurrencies, \textit{Utility tokens} which provide access to an application or service and \textit{Asset tokens} which represent real physical underlying instruments including bonds and derivatives \cite{FINMA:3}.
FINMA classify asset tokens as securities and are therefore subject to the legal requirements and obligations in the jurisdiction they operate in. The Financial Conduct Authority’s (FCA) guidance for cryptoassets defines security tokens as having specific characteristics of a specified instrument like a share or debt instrument and are therefore within the perimeter of the regulator \cite{FCA:4}\cite{CP193:5}. 
The U.S Securities and Exchange Commission (SEC) has also published guidance based on applying the Howey test to a tokenized asset to identify if it falls under Federal securities law \cite{SEC:6}. 
In this paper, we survey efforts to achieve interoperability between distributed ledger technologies and smart contract language support for portability.
We identify factors that enable portability and propose a model that can be used to help assess the maturity of the landscape over time. Lastly, we identify areas for further research that are needed to reach the highest maturity level.

\section{\label{sec:survey}Portability Design and Language Support}
\subsection{\label{sec:definition}Definitions}
We refer to a \textit{tokenized asset as an asset that is considered a security falling under the perimeter of regulation} rather than a cryptocurrency or utility token although the concerns for these are similar.
We define \textit{token portability as the ability to operationally move a token representing all or part of an asset from one distributed ledger to another distinct distributed ledger whilst maintaining history and all legal constraints required.} 
The operation is atomic, in that, the transaction either successfully completes or fails, it cannot be left in an inconsistent state. 
To move an asset from one system to another it is important there is an agreed immutable representation of the asset that includes the rules and events acceptable for the particular asset. 
Any distributed ledger infrastructure employed by existing financial service companies is likely to be permissioned and we consider the factors with this in mind. However, in the future to open the market to a variety of investors connectivity with permissionless distributed ledgers will also be needed. 
For convenience we use the terms distributed ledger, chain and blockchain interchangeably.
Figure 1 depicts portability as the ability to move all or some part of a tokenized asset represented distributed ledger to another distinct permissioned distributed ledger whilst maintaining history and all legal and regulatory obligations.

\begin{figure}[!htb]
    \center{\includegraphics[scale=1]
    {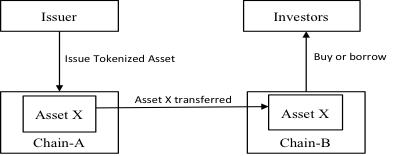}}
    \caption{\label{fig:my-label}Portable Tokenized Asset}
  \end{figure}

\subsection{\label{sec:interoparchs}Architectures for DLT Interoperability}
Interoperability between distributed ledger technologies enables a number of potential use cases, including portable assets, payment-versus-payment, payment-versus-delivery, cross-chain oracles\footnote[1]{\textit{Oracles are trusted intermediates between on and off chain events and data. For more information see https://cryptobriefing.com/what-is-blockchain-oracle/ }}, asset encumbrance and general cross chain contracts \cite{RTHREE:7}.  
For a token to be portable between two ledgers, the ledgers must be interoperable at some level, a search reveals there are currently five high level blockchain interoperability architectures actively under development: 
\begin{enumerate}
\item \textit{Notary schemes} where a group of parties agree to carry out an action on the second blockchain when an event on the first blockchain occurs \cite{RTHREE:7}. 
\item \textit{Side-chains/relay schemes} where components within one blockchain can validate and read events or state in the other blockchain \cite{RTHREE:7}.  
\item \textit{Bridge mechanisms} that are similar to relays where either a smart contract or a software bridge facilitates the communication between blockchains, such as the bridge elements found in Polkadot \cite{POLKA:8}. 
\item \textit{Hash-locking schemes} where operations on two separate blockchains share the same trigger that initates a timed lock during which a transaction completes \cite{RTHREE:7}.
\item \textit{Eventual consistency protocol} proposals such as DeXTT, that transfer tokens existing on one blockchain but are tradeable on multiple blockchains \cite{DEXTT:9}. 
\end{enumerate}

Notary and Relay approaches provide the best support for cross-chain asset portability \cite{RTHREE:7}. 
At present there is no uniform way to transfer tokens representing different assets between different blockchain technologies. 
However, there have been interoperability efforts for cryptocurrencies with Metronome recently announcing interoperability with the MET currency token which is capable of being transferred from Classic Ethereum to Mainnet Ethereum. 
Future plans include support for further interoperability between other blockchain technologies \cite{METRO:10}. 
BTC Relay implements a relay layer between the Bitcoin and Ethereum blockchains that validates Bitcoin transactions in Ethereum smart contracts allowing Bitcoin to pay for Ethereum decentralized applications (dApps) \cite{BTC:11}. 
Keons and Poll identified twelve properties that can be used to evaluate interoperability solutions, all of the properties are relevant to portability to some extent \cite{KEONS:12}. 
However, three key properties directly applicable are semantic interoperability: the agreement on the interpretation of state, syntactic interoperability: ability to communicate through known data formats and protocols and regulation: ensuring compliance with regulation within their operating jurisdiction.
There is a plethora of blockchain technologies in the domain with ongoing research to compare the different interoperability techniques and architectures with no single standard emerging as yet \cite{JOHNSON:13}.  

\subsection{\label{sec:smartlang}Smart Contract Languages}
Tokenized assets are typically represented by smart contracts, for example, on the Ethereum blockchain the ERC20 token standard has been successful in providing a standardized approach for utility tokens with on-going work to define a security token standard ERC1400 \cite{SECTOKEN:14}\cite{ERC:15}. 
ERC token standards are primarily for Ethereum smart contract languages that compile into bytecode stored on each Ethereum node and run on the Ethereum Virtual Machine (EVM) \cite{ETH:16}. 
\\
\\
Solidity, a Turing-Complete statically typed objected oriented language is the most widely used on Ethereum \cite{SOLID:17}. 
Solidity supports complex data types comprised of simple data types that a programmer uses to represent an asset, there is no implicitly enforced representation of an asset or “asset type” within the language specification. 
\\
\\
Flint, a new statically typed smart contract language that compiles into bytecode and runs on the EVM has proposed an asset trait type that represents some asset such as currency and implements behaviors such atomic operations for transfer of part or all of the asset between accounts and prevents non-desirable behavior such as asset destruction \cite{FLINT:18}. 
Use of functions and events that are defined in the trait are enforced by the Flint compiler helping to protect code from attacks such as reentrancy and double spend \cite{ASSETTRAIT:19}. 
Flint is at an early stage of development; the traits constructs are only currently proposals. 
\\
\\
The R3 Corda blockchain uses a modified Java Virtual Machine (JVM) to ensure that contracts written in Java or Kotlin are deterministic. Corda has a library of data types used in financial transactions to provide a common language for smart contract creation \cite{CORDA:20}. 
However, in common with current smart contract languages, rules for the use of the data types and state changes are implemented by the programmer and are not enforced by the type system and compiler. 
\\
\\
DAML is a chain agnostic smart contract language derived from Haskell, that contains native data types to identify parties involved in the contract and control the actions they are allowed to perform; assets are represented using general purpose data structures \cite{DAML:21}. 
\\
\\
Move, the programming language for the Libra blockchain implements assets as first-class resource types that cannot be copied or destroyed and that can only be moved between storage locations. 
The resource types are treated the same as any other type within the language with the safety enforced statically by the type system at compile time.
Move has a concept of a module, cited as similar to a smart contract where a programmer can declare and encode the rules for resource types that are enforced by the type system, Libra coin is implemented in this structure \cite{MOVE:22}.
Flint and Move are not cross-ledger languages; however, first class resource types are potentially a suitable design to enforce standards and protection at compile time to help enable safe cross-chain portability.
\\
\\
A smart contract of a regulated asset is essentially a digital representation of a legal agreement \cite{BRAINE:23}.  
There have been efforts within the financial industry to standardize a common digital representation for derivatives. 
The International Swaps and Derivatives Association (ISDA) have developed a Common Domain Model (CDM) that defines data attributes, trade events and actions in a machine readable and executable format \cite{CDM:24}. 
The derivative data model and lifecycle events defined in the CDM have been implemented in DAML and R3 Corda \cite{EVENTSPEC:25} \cite{HACKATHON:26}. 
The Token Taxonomy Framework (TTF) as an initiatve to define a common language to be used by business, technical and regulatory participants to describe token behaviors and properties. 
The taxonomy is technology neutral and does not mandate programming language or distributed ledger. TTF is not a regulatory framework and currently does not define financial instruments, 
although the base taxonomy could be built upon \cite{TOKENTAXINIT:27}.
\\
\section{\label{sec:survey}Factors Enabling Tokenized Asset Portability}
The structure of today's capital markets is complex environment of issuers that consist of corporations, institutions, governments and multi-lateral entities. Investors and asset managers who consist of insurers, hedge funds, individuals, pension funds, governments, central banks and foundations.
Financial intermediaries such broker-dealers, banks and market infrastructure who act in-between issuers and investors. The market is supported by infrastructure and information providers such as data providers, trade repositories, rating agencies and trading technology platforms. 
The market operates under the protection of regulation and legislation and within social and macroeconomic polices such as tax regimes. \cite{markets:28}
\\
\\
\\
\\
\textbf{Equivalent Market Protectors}. For a tokenized asset to be portable it must operate within the legal rules and protection of the market, that is, the infrastructure supporting the tokenization market must have the equivalent of todays market protecting functions 
including but not limited to Anti-Money Laundering (AML), Know Your Client (KYC), Custodians, Tax Regimes and Regulation. 
A potential solution to this issue is to encode rules in smart contracts agreed between regulators, for example, a Know Your Client smart contract used to check that the identity of buyer is not from a country under sanction. 
An agreed taxonomy that maps between regulators and one that enables a common interpretation of rules is needed. 
\\
\\
\textbf{Technical Features}. Token standards such as ERC1400 are useful to define a common interface, however they do not intrinsically protect against misuse but can be modified to do so.
For example, ConsenSys Codefi Asset representation uses the ERC1400 function \textit{transferWithData}, purposing the data parameter of the function to pass in a
cryptographic certificate generated by the issuer. 
The certificate contains a function identifier to ensure that the certificate can only be used with the particular function and only once.
This can be used in situations where identity is needed such as KYC, where the certificate is issued by a KYC authority only for verified counterparties \cite{GITERC:29}.
However, the implementation of the asset is the responsibility of the developer and the lifecycle or rules of the asset are not enforced by the type system of the language, i.e. the language has no knowledge of the particular asset and it's rules.
Inspired by the language proposals in Flint and Move we believe elevating the representation of the asset in to the language type system allows the compiler to enforce rules and constraints at compile time helping with standardization and code safety.
Interoperability is also supported as the compiled representation of the asset aids semantic and syntactic interoperability thus aiding standardization across different ledger technologies.

\section{\label{sec:survey}Portability Maturity Model}

Our research indicates that there are three factors affecting the successful portability of tokenized assets:

\begin{enumerate}
 \item \textit{Chain Interoperability}. With two sub properties: \textit{Semantic interoperability} that enables full agreement on the interpretation of state. 
  That is, state transferred from chain A to chain B is not transformed in such a way that properties such as history, identity and ownership are destroyed. 
  At the highest level of maturity this includes permissioned to permissionless transfers where consensus mechanisms are different, for example, notary consensus in one chain and proof of stake consensus in another.
  \textit{Syntactic interoperability} that provides cross-ledger shared data formats and protocols of heterogeneous technologies.

  \item \textit{Smart Contract Programming Language Asset Support}. The ability of a smart contract programming language to be compatible with an idealized homogeneous smart contract language with a type system enforced representation of assets. 
  The language we describe here is idealized, in reality a number of languages would be expected to implement to the same asset representation standard. 
  We expect the idealized language to be homogeneous but deployable to heterogeneous infrastructures that adhere to factors one and three. 

  \item \textit{Equivalent Market Protectors}. 
  Equivalent market protectors and infrastructure providers are needed to ensure tokenized assets comply with regulatory, legal and compliance requirements.
  Today, market processes are off-chain and in some cases accessible by API. 
  To support a fully functioning distributed ledger based market ecosystem inclusive of portability, where possible protecting functions will need to be on-chain represented by robust smart contracts. 
  This will allow greater automation, reduce operational dependancy of external services and confidence in agreed behaviours.
  As we have already indicated these services include but are not limited to Know Your Client (KYC), Anti-Money Laundering (AML), Custody, Tax Regimes and Regulatory Obligations such as reporting.
  This is complicated when issuance is limited to one distributed ledger within a single jurisdication with complexity increasing across ledgers that may be operating in differing jurisdictions.
  A common regulatory taxonomy is needed define equivalence functions between jurisdictions. 
\end{enumerate}

In table one we list the three factors and three increasing levels of maturity to reach a fully functioning tokenized asset ecosystem that can support portability.

\section{\label{sec:survey}Conclusion}

In this paper we have surveyed smart contract languages and distributed ledger interoperability architectures to consider their ability to support portability of tokenized assets.
The large number of differing technologies still emerging in this fast-moving space has the potential to make interoperability and standardization difficult. 
Which in turn may lead to the fragmentation of an already complex market and fail to unlock the cited benefits of portability.  
We identified that equivalent market protectors will be needed, on-chain for level 3. Full on-chain equivalence has a dependency on a idealized smart contract language that is functionality able to support these requirements. 
Also, in general this is a non-trivial endeavor requiring collaboration and agreement on 
standards across technology, compliance, regulation, legal and market participants. Note: we chose not to include infrastructure providers as a factor within the maturity model our reasoning being that achieving any level 3 factor intrinsically requires a capable provider or vendor.
To help gauge the maturity of current and future technology and supporting functions we have presented a three level maturity model. 
Today's state appears to be at level one with some chain interoperability approaching level 2. 
There is still a great deal of work to be done before level 3 is reached. 

\section{\label{sec:survey}Further Work}

To provide the technology capability needed for level 3 maturity, future work is required to define a type system and rules that represents a real-world asset to understand how this works in practice.
After successful definition we recommend analysis of the type system of existing smart contract languages to gauge if existing languages are truly fit for purpose or a next generation novel language is required.
Interoperability is a key feature for portability and we recommend further research into how agreement of state and standardization can be achieved. 
An idealized standard smart contract language to meet the requirements of equivalent market protectors is a future requirement. We recommend a multi-disciplined approach of law makers and market incumbents to discover the detailed requirements of such a capability both technically and functionally.

\begin{table}[H]
  \caption{\label{tab:my-label}Tokenized Asset Portability Maturity Model}
    \center{\includegraphics[height=16cm, width=12.5cm]{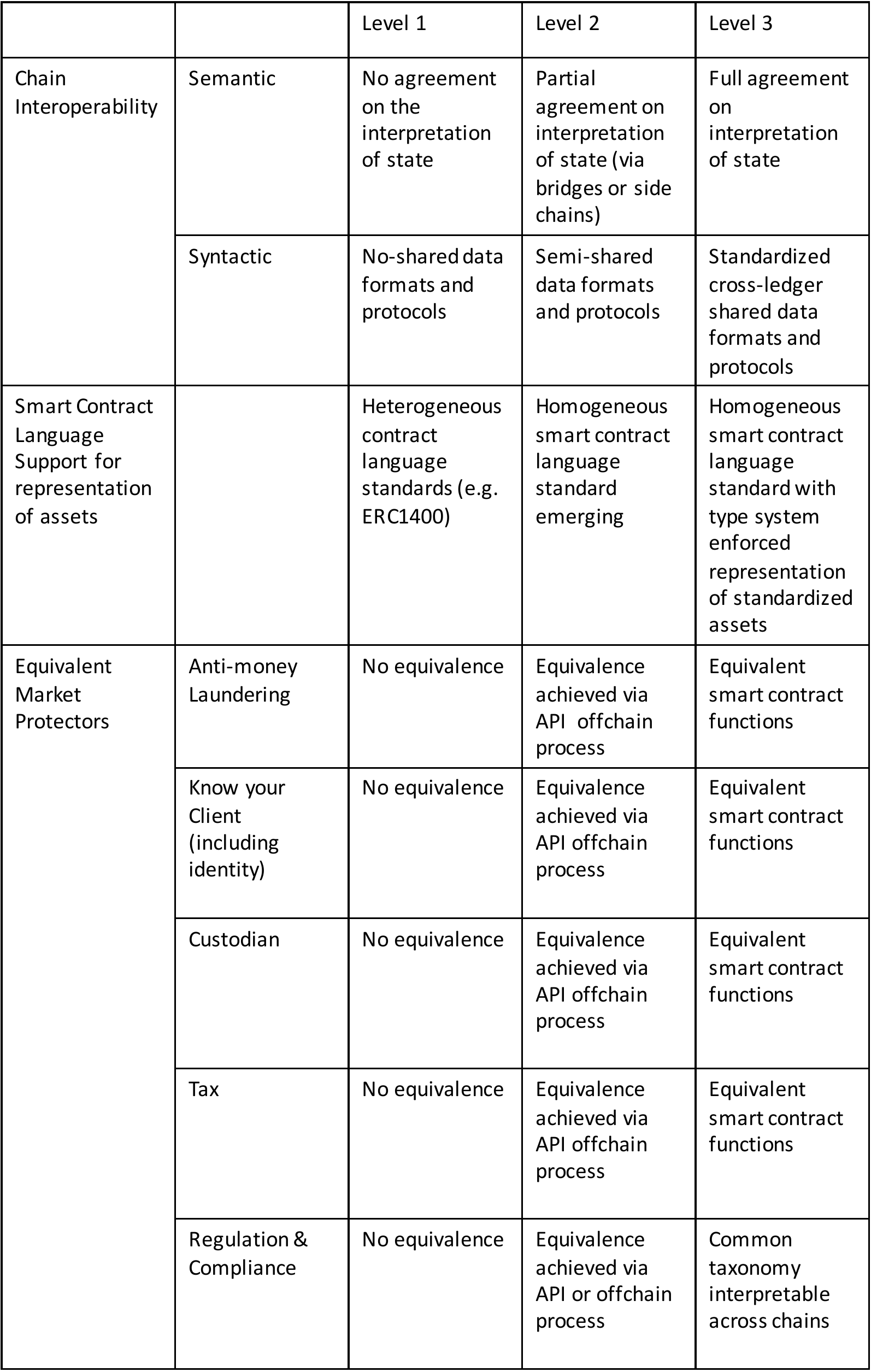}}
\end{table}

\bibliography{Factors_in_the_Portability_of_Tokenized_Assets_on_Distributed_Ledgers}
\bibliographystyle{ieeetr}
\end{document}